Research article

# Compensatory mutations cause excess of antagonistic epistasis in RNA secondary structure folding

Claus O Wilke*[1], Richard E Lenski[2] and Christoph Adami[1,3]

Address: [1]Digital Life Laboratory 136–93, California Institute of Technology, Pasadena CA, 91125, USA, [2]Center for Biological Modeling, and Center for Microbial Ecology, Michigan State University, East Lansing, MI 48824, USA and [3]Jet Propulsion Laboratory 126–347, California Institute of Technology, Pasadena CA 91109, USA

Email: Claus O Wilke* - wilke@caltech.edu; Richard E Lenski - lenski@msu.edu; Christoph Adami - adami@caltech.edu

* Corresponding author

**Background:** The rate at which fitness declines as an organism's genome accumulates random mutations is an important variable in several evolutionary theories. At an intuitive level, it might seem natural that random mutations should tend to interact synergistically, such that the rate of mean fitness decline accelerates as the number of random mutations is increased. However, in a number of recent studies, a prevalence of antagonistic epistasis (the tendency of multiple mutations to have a mitigating rather than reinforcing effect) has been observed.

**Results:** We studied *in silico* the net amount and form of epistatic interactions in RNA secondary structure folding by measuring the fraction of neutral mutants as a function of mutational distance $d$. We found a clear prevalence of antagonistic epistasis in RNA secondary structure folding. By relating the fraction of neutral mutants at distance $d$ to the average neutrality at distance $d$, we showed that this prevalence derives from the existence of many compensatory mutations at larger mutational distances.

**Conclusions:** Our findings imply that the average direction of epistasis in simple fitness landscapes is directly related to the density with which fitness peaks are distributed in these landscapes.

# Background

Epistatic interactions between mutations in different genes, or different sites within the same gene, can substantially influence the dynamics of evolving populations. Although adaptation by natural selection depends on occasional beneficial mutations, in fact most mutations are neutral or deleterious. Therefore, if mutations are randomly added to an organism's genome (as opposed to incorporated by natural selection), then fitness tends to decline with mutation number. Mutations whose effects are independent of one another are termed *multiplicative*. If mutations interact so that their combined effect on fitness is greater than expected from their individual effects, then epistasis is said to be *synergistic*. For deleterious mutations, synergistic epistasis means that the decline in average fitness accelerates as more random mutations are added (such epistasis is termed *negative* by Phillips et al. [1]). By contrast, if deleterious mutations interact so that their combined effect is smaller than expected under the multiplicative model, then epistasis is called *antagonistic* (or *positive* in the terminology of Phillips et al.). Antagonistic epistasis therefore implies unexpected robustness to the effects of multiple deleterious mutations [2].

The predicted effects of these various types of mutational interactions on population dynamics are well studied (for reviews, see [3–5]), but in most models the types of interactions are ad hoc because it is not known how well the assumed interactions reflect reality. In particular, the question of what kinds of interactions are typical, and under what conditions, is addressed only rarely.

Recent work has shown that epistatic interactions can be interpreted as curvature in the fitness surface [6], which implies that geometry imposes limitations on the types of epistasis that can be observed for any given genotype. Closely related to this geometric approach is the realization that epistasis must be studied with respect to a particular reference sequence [7–11], and that the forms of epistatic interactions change when the reference is changed. Wilke and Adami [11] proposed that inhomogeneities in the distribution of genotypes give rise to directional epistasis (directional epistasis is the net epistasis averaged over many pairs of mutations) among deleterious mutations, and that the sign of directional epistasis provides information about the density gradient of high-fitness sequences within a cluster. Hence, evolutionary hypotheses that depend crucially on the sign of epistasis, including certain theories on the evolution of recombination [12–15] and on the speed of mutation accumulation in asexual populations [16–18], will also depend on the density distribution of fit genotypes, and on the topology of those regions of genotype space into which populations move under the pressures of mutation and selection.

The argument relating directional epistasis to the distribution of high-fitness sequences within a cluster of fit sequences goes as follows (Figure 1). Consider a sequence that lies near the *center* of a dense cluster of high-fitness sequences. Such a sequence is surrounded by many other high-fitness genotypes, and the average harm done by a single mutation is therefore relatively low. But as more random mutations are added, high-fitness genotypes become less common, and the average harm of multiple mutations is stronger than what the effect of single mutations indicated. This increasingly harmful effect of mutations at greater distances away from the original sequence corresponds to *synergistic* epistasis. By contrast, if a sequence is located at the periphery of a high-fitness cluster, or in a region of low density of high-fitness sequences, then this synergistic tendency is lessened, and we might even observe antagonistic epistasis for this sequence. One prediction of this hypothesis of density inhomogeneities is that the average fitness effect of single mutations must be correlated with the strength and direction of epistatic interactions at larger mutational distances. This prediction was confirmed in RNA secondary structure folding and in digital organisms [11]. A similar effect was observed for quantitative trait loci in mice [19].

While the hypothesis of Wilke and Adami [11] explains some of the variation within the observed range of epistatic interactions, it fails to predict whether we should always expect an excess of synergistic epistasis, or whether antagonistic interactions might dominate in certain situations. Recent experiments indicate that synergistic and antagonistic interactions are both common when pairs of mutations are considered, but that, overall, the two types of interactions either roughly cancel each other out or perhaps even produce an excess of antagonistic interactions [20–29]. Only one study of this type reported an excess of synergistic interactions [30]. Moreover, two simple systems with computational landscapes, self-replicating computer programs [2] and RNA secondary structure folding [11], both exhibited a clear excess of antagonistic interactions.

What factors may explain a tendency toward antagonism? Here are four possible explanations, and there may be others:

1. Biased recovery: In some experimental systems, it is difficult to isolate genotypes with very low fitness, including non-viable ones. By allowing random mutations to accumulate only in surviving lineages, a researcher may inadvertently bound mean fitness away from zero, and thereby bias the shape of the fitness function versus mutation number from synergy toward antagonism.

2. Multiple hits: To the extent that multiple mutations hit the same gene or pathway, subsequent mutations will more likely be neutral than initial ones, because the same function cannot be destroyed twice. In effect, one mutation may convert one or more genes into pseudo-genes, or junk DNA, in which subsequent mutations have no further harmful effect.

3. Intrinsic or evolved form of interaction: The dynamics of metabolic pathways and gene regulation might be inherently structured to produce more antagonistic than synergistic interactions. Even if there is no necessary biochemical tendency in this direction, it is possible that pathways and their regulation have evolved to promote such interactions. While this explanation is vague, a failure to identify any other explanation for observed patterns of mutational interaction would suggest a closer examination of models of metabolic pathways and gene regulation with this issue in mind.

4. Compensatory mutation: Consider a genotype that is on a peak (or plateau) in a fitness landscape, such that all mutations are either deleterious or neutral. As one moves away from this genotype by adding mutations, imagine that the ratio of deleterious to neutral mutations stays constant, as does the average effect of a deleterious mutation. But as the distance away from this peak increases, a portion of the mutant distribution may move into the vicinity of one or more other fitness peaks, separated from the original one by a valley of deleterious mutations. Mutations that lead out of the valley onto a new fitness peak are, of course, beneficial mutations. If a mutation is beneficial because it restores some previously existing structure or function, then we can call it a compensatory mutation. If the frequency of compensatory mutations increases as one moves out from the original peak, this effect will slow down the decay of mean fitness with increasing mutational distance, even without *selection* for compensatory mutations.

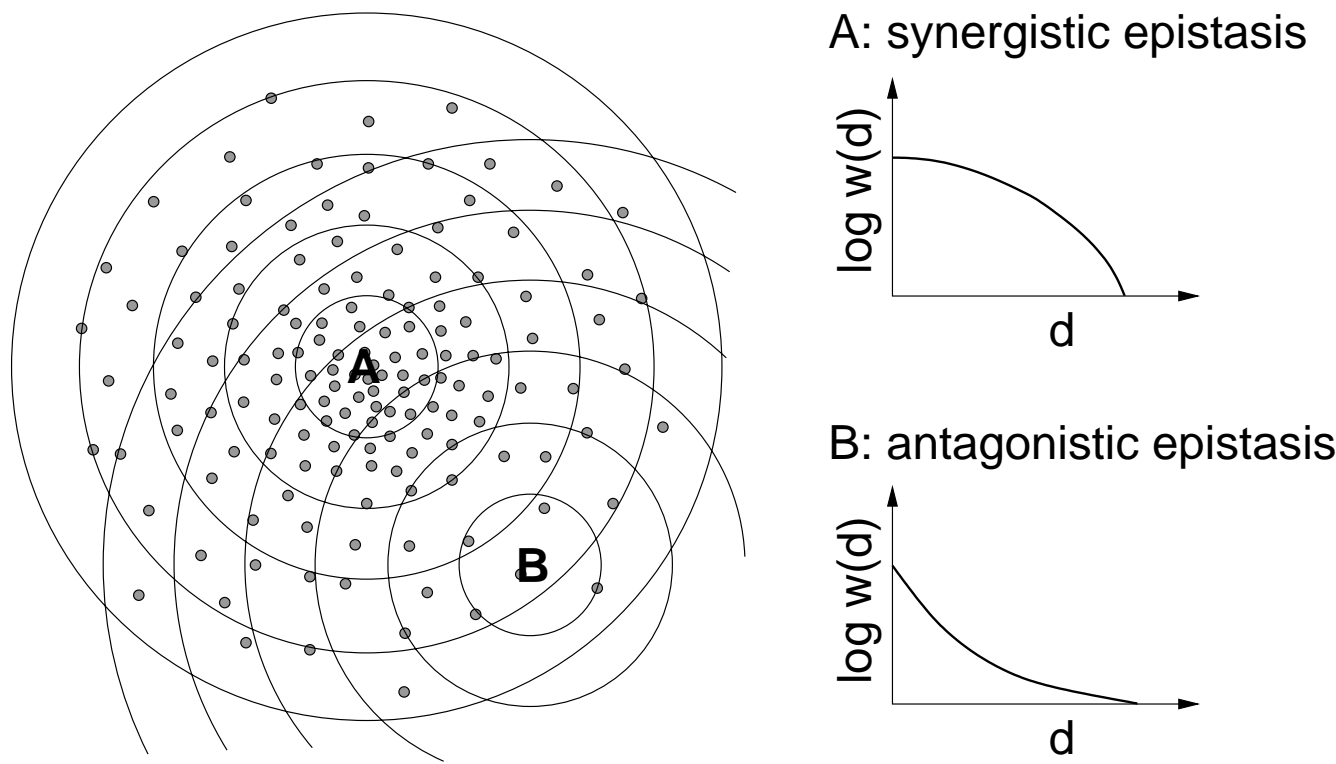

**Figure 1**
Relationship between the distribution of high-fitness sequences and directional epistasis, according to Wilke and Adami [11]. The drawing on the left visualizes genotype space, with the small filled circles representing high-fitness genotypes. A and B are two particular reference sequences, and the concentric rings around A and B indicate the mutants that are a fixed Hamming distance away from either A or B. In the case of A, the average fitness $w(d)$ of the sequences at Hamming distance $d$ from A decays faster at higher $d$ than at lower $d$, and therefore A shows synergistic epistasis. In the case of B, the decay of $w(d)$ slows down as $d$ increases, and hence B shows antagonistic epistasis.

In this paper, we focus on a fairly simple system (RNA secondary structure folding), which avoids the first explanation above because of complete information, and avoids the second and third explanations because there is only a single gene with dichotomous fitness (0 or 1) in the system. Thus, we can focus on whether, and to what extent, compensatory mutations that precisely restore the previously existing folding structure contribute to an excess of antagonistic interactions.

In order to analyze the topology and structure of fitness landscapes, and assess their effect on epistasis, we need a system where each sequence in a well-defined space is assigned a unique fitness that is independent of the frequency of other genotypes in the population. RNA secondary structure folding is a well-studied model system. In this system, the secondary structure of an RNA molecule is taken as a predictor of the fitness of the underlying sequence [31–36]. RNA folding has the advantage that all genotypes can be separated naturally into only two classes, the viable ones (those that fold into a particular target structure) with fitness 1, and the non-viable ones (those that fail to fold into that exact structure), which are assigned fitness 0. Such a classification permits a precise mathematical analysis, while at the same time many important properties of high-dimensional fitness landscapes are retained [37]. The set of all viable sequences can be decomposed into smaller sets called neutral networks [38,39]. Two viable sequences belong to the same neutral network if one sequence can be transformed into the other through a series of single point mutations, without passing through a sequence with lower fitness along the way. If no such series of mutations exists, then the two viable sequences belong to different neutral networks.

As discussed in previous publications [2,11,25], the average fitness $w(d)$ at a Hamming distance $d$ from a particular reference sequence can be used for studying directional epistasis in a large number of genotypes. The value of $w(d)$ is obtained by averaging over the individual fitnesses of all sequences that can be obtained from the reference sequence by introducing $d$ mutations. By definition, $w(0)$ is the fitness of the reference sequence, and we have always $w(0) = 1$. The value of $w(1)$ is given by 1 minus the mean effect of single mutations and, in general, the value of $w(d)$ is given by 1 minus the mean effect of $d$ mutations. Moreover, since we assume in this study that all individual fitnesses are either 0 or 1, $w(d)$ corresponds also to the fraction of viable sequences at distance $d$ from the reference sequence. We will make use of this property repeatedly throughout this paper.

We expect that $w(d)$ decreases roughly exponentially as $d$ increases, for the following reason. If the sequences are sufficiently long that any back mutations can be neglected, then genotype space can be visualized as a tree (Fig. 2). If the number of neutral neighbors (the neutrality, $\nu$) of all viable sequences is approximately the same, and the fraction of neutral mutations of a viable sequence is on average much larger than the probability that a completely random sequence is viable, then $w(d + 1)$ can be approximated by $[\nu/(L\kappa-L)]w(d)$, where $L$ is sequence length and $\kappa$ denotes the number of different bases that can occur ($\kappa = 4$ for RNA). Thus, we can write $w(d) = \exp(-\alpha\, d)$ with $\alpha = 1 - \nu/(L\kappa-L)$. The condition that the fraction of neutral mutants $\nu/(L\kappa-L)$ is much larger than the total fraction of viable mutants among all possible sequences is crucial for the exponential decay of $w(d)$. Consider by contrast the Russian Roulette model by Gavrilets and Gravner [40], in which all sequences are randomly assigned a fitness value of either 0 or 1. The total fraction $p$ of viable sequences in this model is equal to the expected fraction of neutral mutants of any sequence with fitness 1, and $w(d) \approx p$ for $d > 0$ in this model.

In the general case, $\nu$ is not constant. If $\nu$ changes systematically with increasing $d$, then $w(d)$ will decay slightly faster or slower than a perfect exponential. We incorporate this by writing [2]

$$w(d) = \exp(-\alpha d^\beta). \quad (1)$$

The parameter $\beta$ represents the *deviation* from the exponential expectation. Because an exact exponential decay means that the average effect of a new mutation is independent of how many mutations have accumulated previously, $\beta$ is a measure of the net effect of epistatic interactions. In the following, we refer to this net effect as directional epistasis, and to $\beta$ as the epistasis parameter. $\beta > 1$ corresponds to synergistic epistasis, whereas $\beta < 1$ corresponds to antagonistic interactions.

Consider the fitness landscape depicted in Figure 2, in which circles represent viable sequences while crosses are non-viable ones. This particular landscape contains two separate neutral networks, of which one (indicated by bold solid lines) is connected to the reference sequence at distance $d = 0$. At $d = 2$ in the upper branch, a compensatory mutation leads onto a second neutral network (shown as bold dashed lines). For such "fitness trees" anchored at particular reference sequences, we can measure both the decline in average fitness with mutational distance, $w(d)$, as well as the change in the average neutrality as we increase $d$. The average neutrality at distance $d$ simply reflects the connectedness of the neutral network at that distance. Average neutrality at mutational distance $d$ is defined as the average number of neutral neighbors of a sequence that is separated by $d$ mutations from the reference sequence and that is part of the neutral network under investigation. As per the arguments by Wilke and Adami [11], we expect the connectedness to decrease with $d$ as we move away from a reference sequence in the center of a high-fitness cluster. By contrast, we expect connectedness to increase (or at least decrease more slowly) if the reference sequence is on the fringe of such a cluster. Figure 3 shows $w(d)$ and average neutrality for examples of "fringe" and "center" reference sequences. The fringe sequence (Fig. 3a) has strongly antagonistic net epistasis, and neutrality increases with distance. By contrast, the center sequence (Fig. 3b) shows a slightly synergistic tendency and declining neutrality with distance. This association agrees with our expectation that more antagonistic curves $w(d)$ are caused by an increase in the density of neutral sequences with $d$, whereas more synergistic curves result from a corresponding decline.

## Results

For 100 randomly chosen RNA sequences of length $L = 76$, we measured both the fraction of neutral mutants $w(d)$ as well as the average neutrality of neighbors of neutral mutants as a function of the distance $d$. We determined $\alpha$ and $\beta$ from fitting $\exp(-\alpha d^\beta)$ to $w(d)$. Moreover, we performed a linear regression on the average neutrality to quantify the change in neutrality with distance from the reference sequence, for each of the random sequences serving as a reference. Among the 100 reference sequences that we studied, the change in neutrality was sufficiently linear to make the slope $m$ an excellent indicator of the general trend (increasing or decreasing neutrality with distance $d$), even if the true relationship was not always precisely linear.

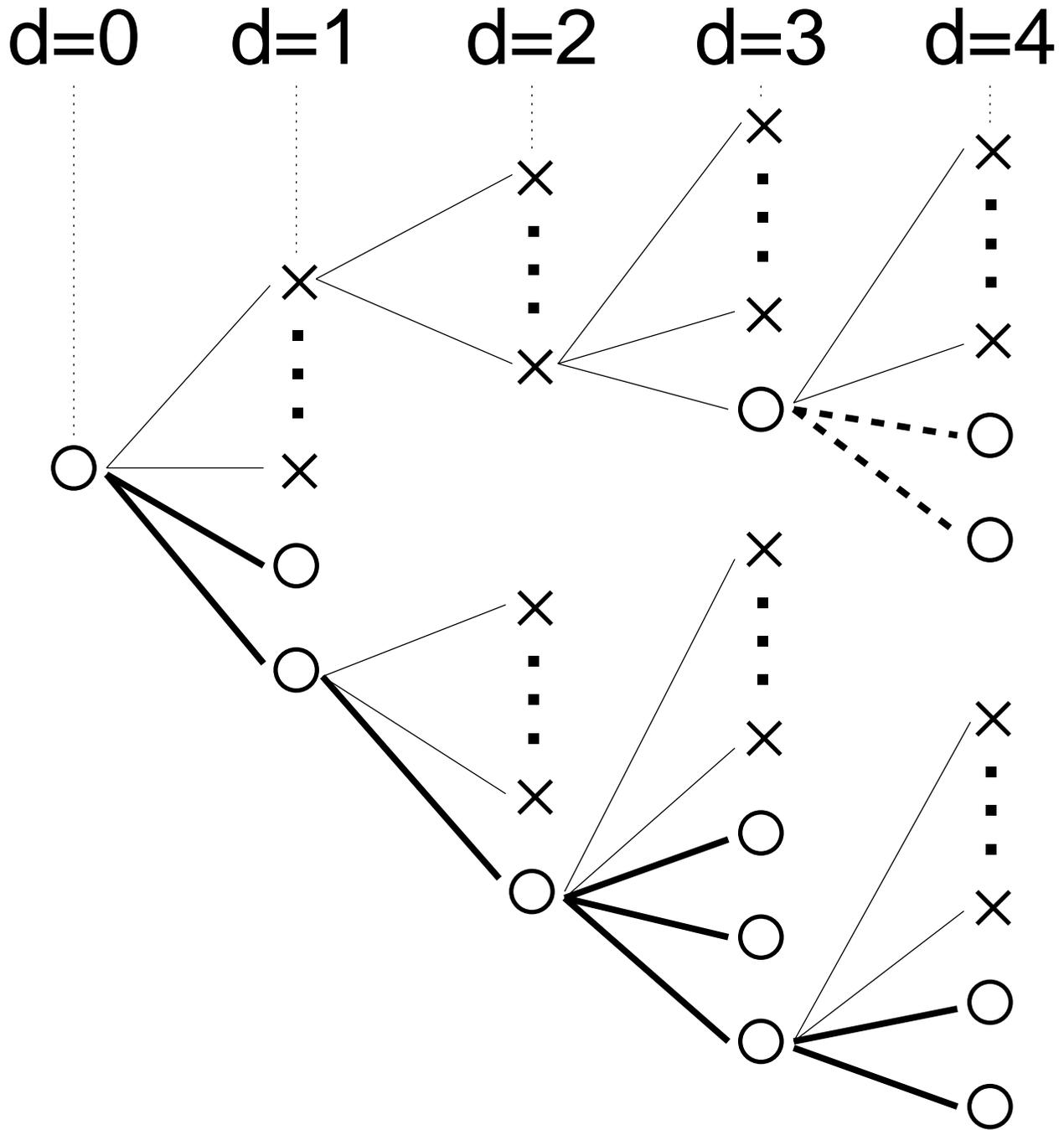

**Figure 2**
Schematic drawing of a fitness landscape. Circles represent viable sequences, and crosses represent non-viable ones. The sequence at $d$ = 0 serves as the reference sequence. It has two viable and a number of non-viable mutational neighbors. The viable mutants have further viable and non-viable mutational neighbors, and so on. All viable mutants on the lower branch form a single neutral network. On the upper branch, a new neutral network emerges at a mutational distance $d$ = 3 from the reference sequence.

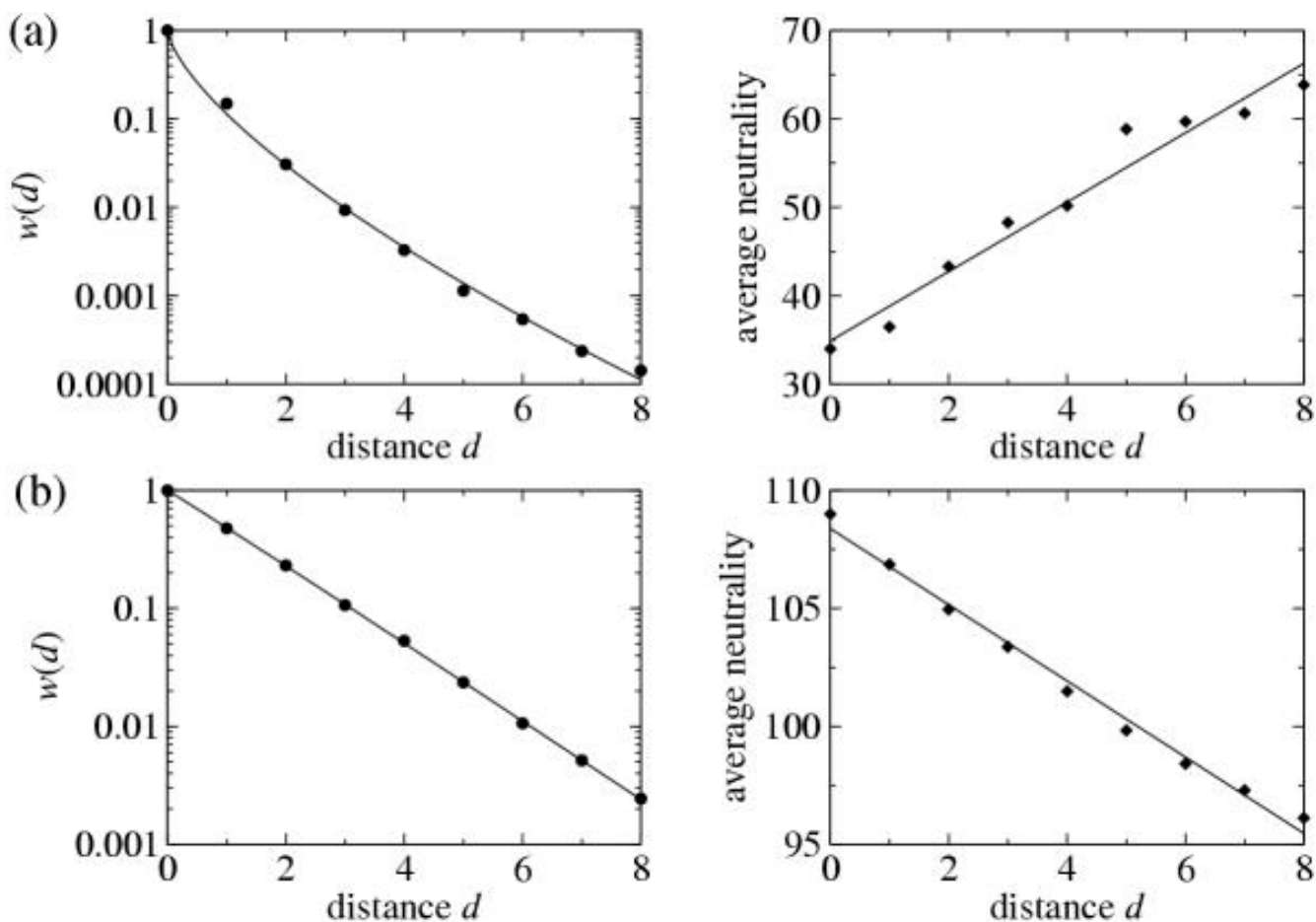

**Figure 3**
Function $w(d)$ (left panel) and average neutrality (right panel) at distance $d$, for two reference sequences. Points are measurements, and lines represent best fits [function $\exp(-\alpha d^\beta)$ for $w(d)$, and function $md + n$ for average neutrality]. (a): A case of strongly antagonistic epistasis ($\beta = 0.688$), which is associated with increasing neutrality with $d$ (right panel). (b): A more synergistic case ($\beta = 1.017$), and a corresponding decline in neutrality with $d$.

Figure 4 shows this measured slope $m$ versus the fitted epistasis parameter $\beta$ for each of the 100 reference sequences. The two quantities are very strongly correlated (correlation coefficient $r = -0.960$, $p < 0.0001$). However, the distributions of $m$ and $\beta$ differ in an important respect. The values of $m$ are distributed roughly symmetrically around zero, meaning that the average neutrality increases as often as it decreases with distance from the reference sequence. The epistasis parameter $\beta$, on the other hand, lies almost entirely below unity. Even when the average neutrality decays quickly as one moves away from the reference sequence, the corresponding value of $\beta$ indicates at most a very slight tendency toward synergistic epistasis. This is precisely the pattern of excess antagonistic epistasis that demands an explanation.

If compensatory mutations were responsible for the shift of $\beta$ to values below one, then by adjusting the function $w(d)$ to *remove* their contribution we should find a distribution of $\beta$ symmetrical around one. As described in the

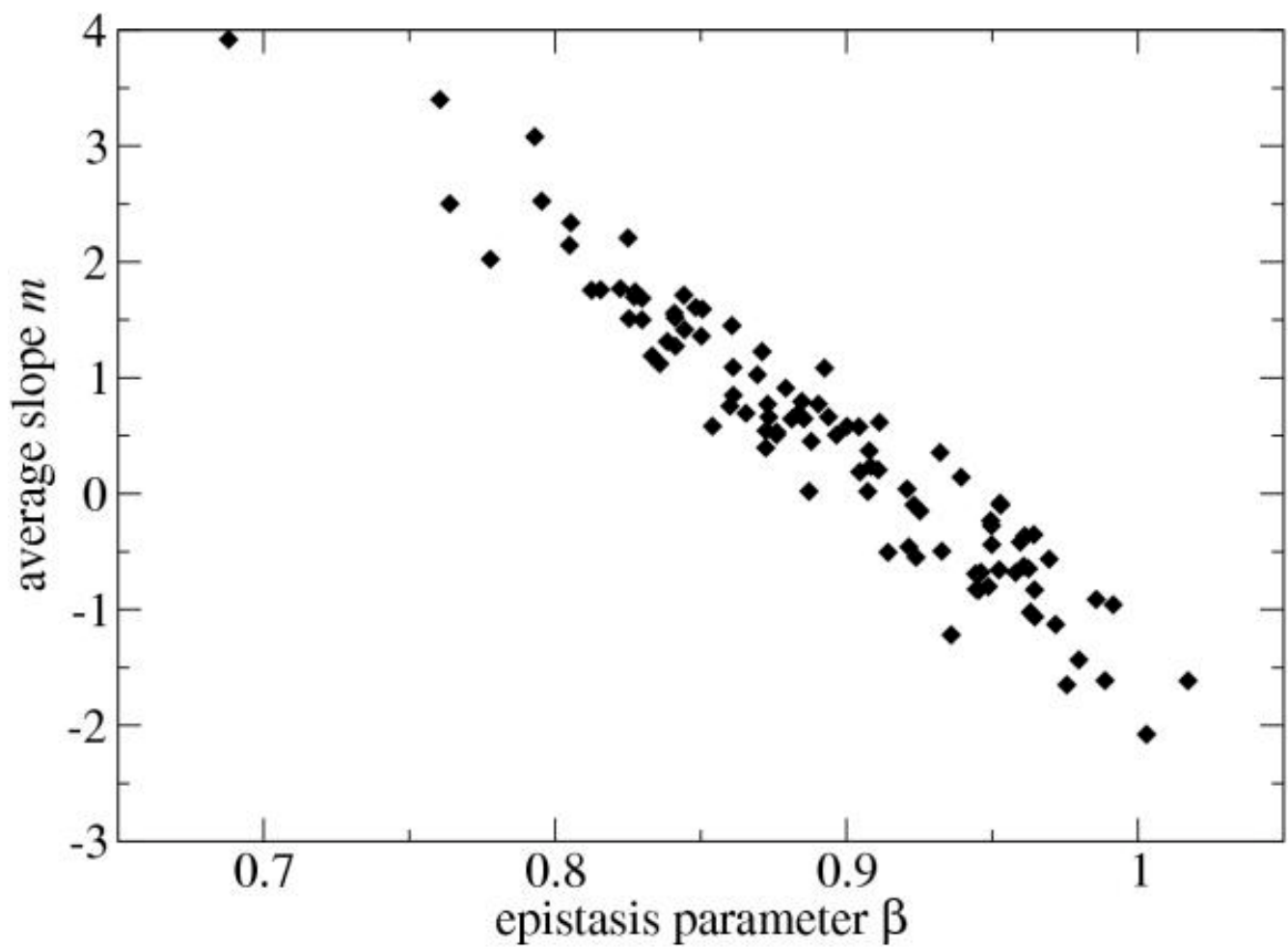

**Figure 4**
Correlation between the average change in neutrality, *m*, and the epistasis parameter, β, for 100 reference RNA sequences. The negative correlation is highly significant ($r = -0.960$, $p < 0.0001$).

Methods section, it is possible to separate the contribution to $w(d)$ of neutral mutations that lie on the same network $w_{neut}(d)$ from the contribution $w_{comp}(d)$ that reflects compensatory mutations from neighboring neutral networks onto the reference network. Figure 5 shows $w(d)$, $w_{neut}(d)$ and $w_{comp}(d)$, estimated using method M2, for a typical example. The contribution of $w_{neut}(d)$ to $w(d)$ dwindles as the mutational distance $d$ increases, while $w_{comp}(d)$ becomes increasingly dominant. In other words, at large distances $d$, most of the viable sequences arise through compensatory mutations.

After partitioning the contributions of $w_{neut}(d)$ and $w_{comp}(d)$ for all 100 functions $w(d)$, we determined α and β for the functions $w_{neut}(d)$ in the same way as was done for $w(d)$. The functions $w_{neut}(d)$ can be thought of as the decay functions without "contamination" by compensatory mutations. Thus, the differences in α and β for $w(d)$ and $w_{neut}(d)$ must be ascribed to compensatory mutations. Figure 6 shows α versus β for both $w(d)$, shown as triangles, and $w_{neut}(d)$, calculated by method M2, shown as circles. The removal of compensatory mutations causes a pronounced upward shift of β toward more synergistic epistasis. The average decay parameter α is largely unaffected. For $w(d)$, average <β> = 0.893 ± 0.006 (standard error of the mean), while for $w_{neut}(d)$ the average <β> = 0.983 ± 0.004 by M2 or 1.046 ± 0.005 by M1. Figure 7 shows that the removal of compensatory mutations yields a distribution of β that is approximately symmetric around one, in contrast to the pronounced excess of antagonistic epistasis (β < 1) when compensatory mutations are included.

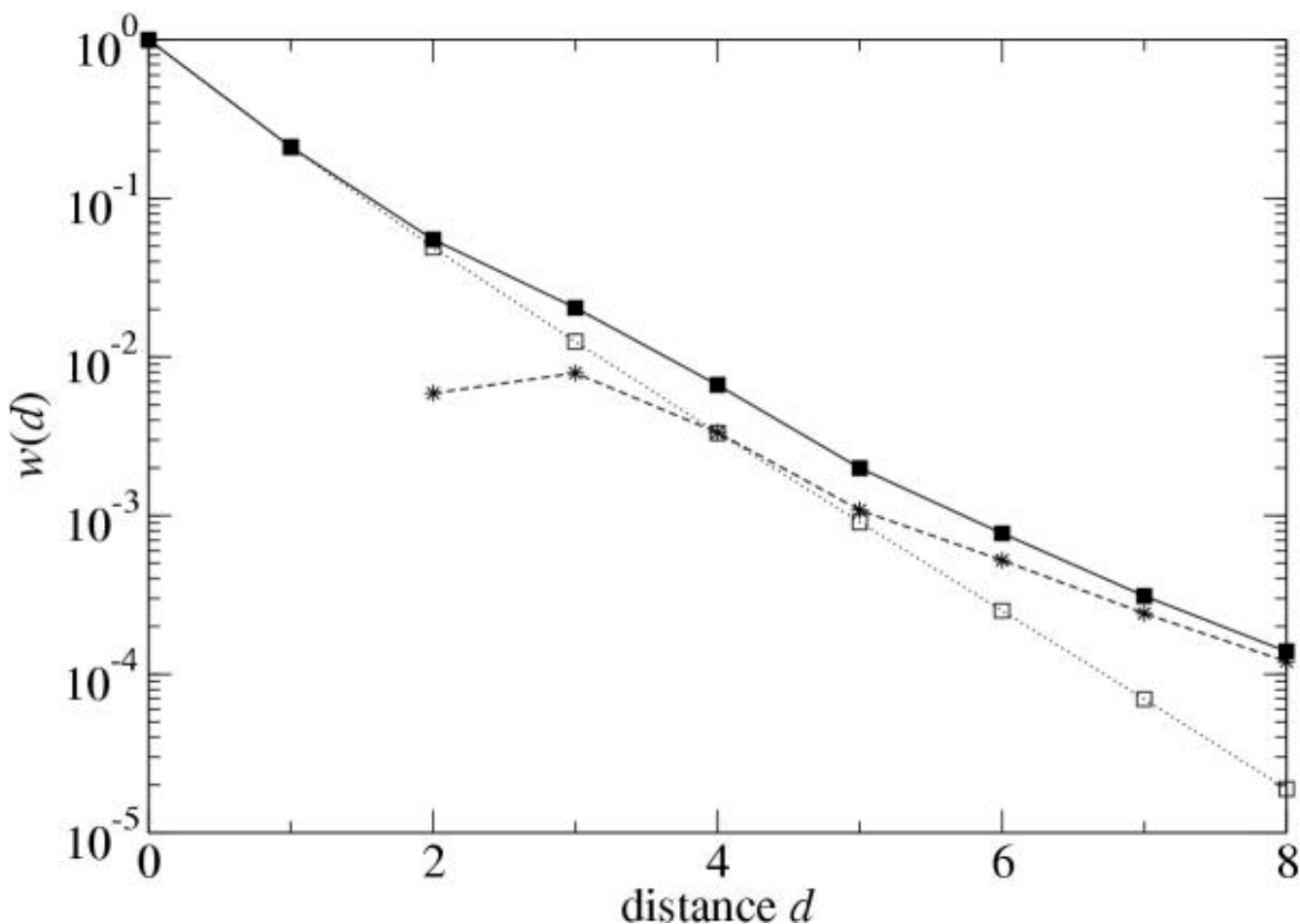

**Figure 5**
Functions $w(d)$ (■), $w_{neut}(d)$ (□), and $w_{comp}(d)$ (*), calculated using method M2, for a representative sequence. The lines merely connect the points to guide the eye.

**Discussion**
The correlation between average mutational effects ($\alpha$) and directional epistasis ($\beta$) described by Wilke and Adami [11] implies that whether a reference sequence shows synergistic or antagonistic epistasis depends on whether it lies near the center of a dense cluster of high-fitness sequences in genetic space, or on the fringes of such a cluster. Hansen and Wagner [10] independently suggested that the relative frequency of synergistic and antagonistic interactions depends on the distance of the genotype in question from the optimum, and thus from the center of the cluster.

In this study, we demonstrated a very strong correlation between $\beta$, which governs the net direction of epistasis, and the mean change in neutrality per added mutation away from the reference sequence, $m$ (Fig. 4). This correlation is fully consistent with the hypothesis that directional epistasis depends on the location of the reference sequence relative to other high-fitness genotypes. Yet, this correlation does not explain the overall preponderance of antagonistic epistasis observed in two computational systems [2,11], nor the absence of overall synergistic epistasis in most biological systems that have been carefully studied in this regard. Although we found wide variation in the epistasis parameter for RNA secondary structure folding (Fig. 4), 98 of 100 reference sequences showed an excess of antagonistic epistasis ($\beta < 1$) and only 2 exhibited more synergism ($\beta > 1$). We demonstrated that this overall excess of antagonistic interactions results from compensatory mutations (Figs. 5, 6, 7), which are evidently important for the shape of the fitness function whether or not a sequence is in the middle of a dense cluster.

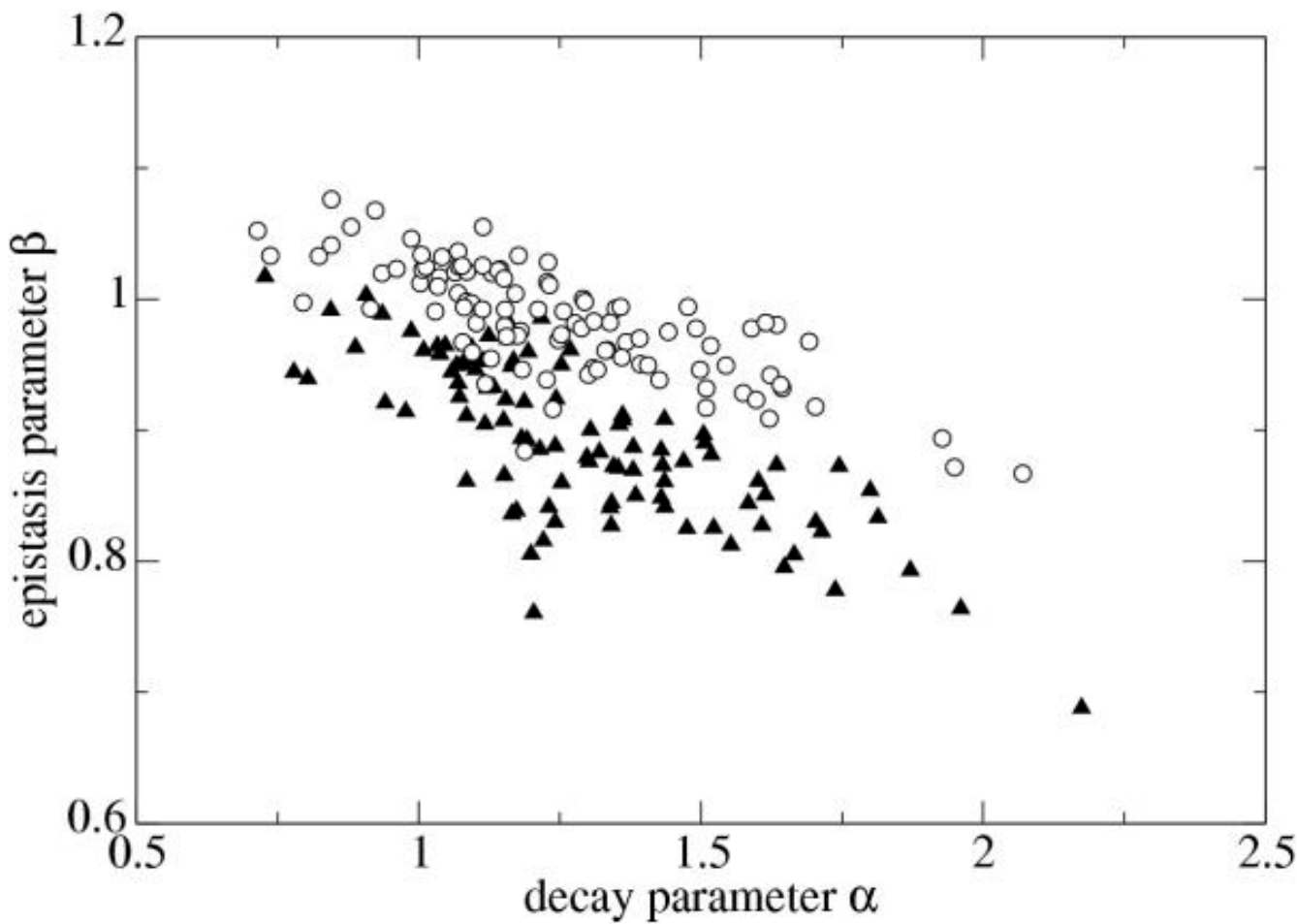

**Figure 6**
Epistasis parameter β and decay parameter α for 100 reference RNA sequences, plotted separately for the full fitness function $w(d)$ (triangles) and for the purely neutral contribution $w_{neut}(d)$ (circles, method M2).

The trajectory for mean fitness as random mutations accumulate is generally one of decay because many more mutations are deleterious than are beneficial. The decay function is therefore sometimes loosely discussed as if it depended only on deleterious mutations (or only deleterious and neutral mutations). This focus on declining fitness tends to ignore compensatory mutations because they are beneficial, albeit conditionally. But compensatory mutations cannot be ignored if their frequency varies systematically in genetic space. In particular, if compensatory mutations are an increasing fraction of all viable mutations as the distribution moves away from a local fitness peak, then they will affect the shape of the decay function (Fig. 5) and the resulting epistasis parameter (Figs. 6, 7).

In the present paper, we have used the term "compensatory mutation" for all mutations that are not on the main neutral network, irrespective of their distance to this network. This is a broad definition of compensatory mutations, because it encompasses not only those mutations that directly counteract a particular deleterious mutation, but also mutations that establish a new neutral network far away from the main network. The latter type of mutations could also be called advantageous mutations. Here, we have lumped both types together, because in this way the presentation of our results is less cluttered, and the mathematics of separating $w(d)$ into the various contributions is easier to follow, than if we were distinguishing three different types of mutations. An additional argument for calling all these mutations compensatory is that they can produce a fitness gain only in association with one or more mutations that would otherwise be deleterious, that is, none of the mutations that could be termed advantageous is accessible directly from the main neutral network. On the other hand, our definition excludes mutations that restore a genotype to the main neutral network (even mutations that are not reversions in the strict sense). Again, we chose this definition for mathematical precision and clarity. The definition is not as important as the substantive conclusion: The fact that it is possible to restore a previously existing structure by many non-trivial mutational pathways is sufficient to cause pronounced antagonistic directional epistasis.

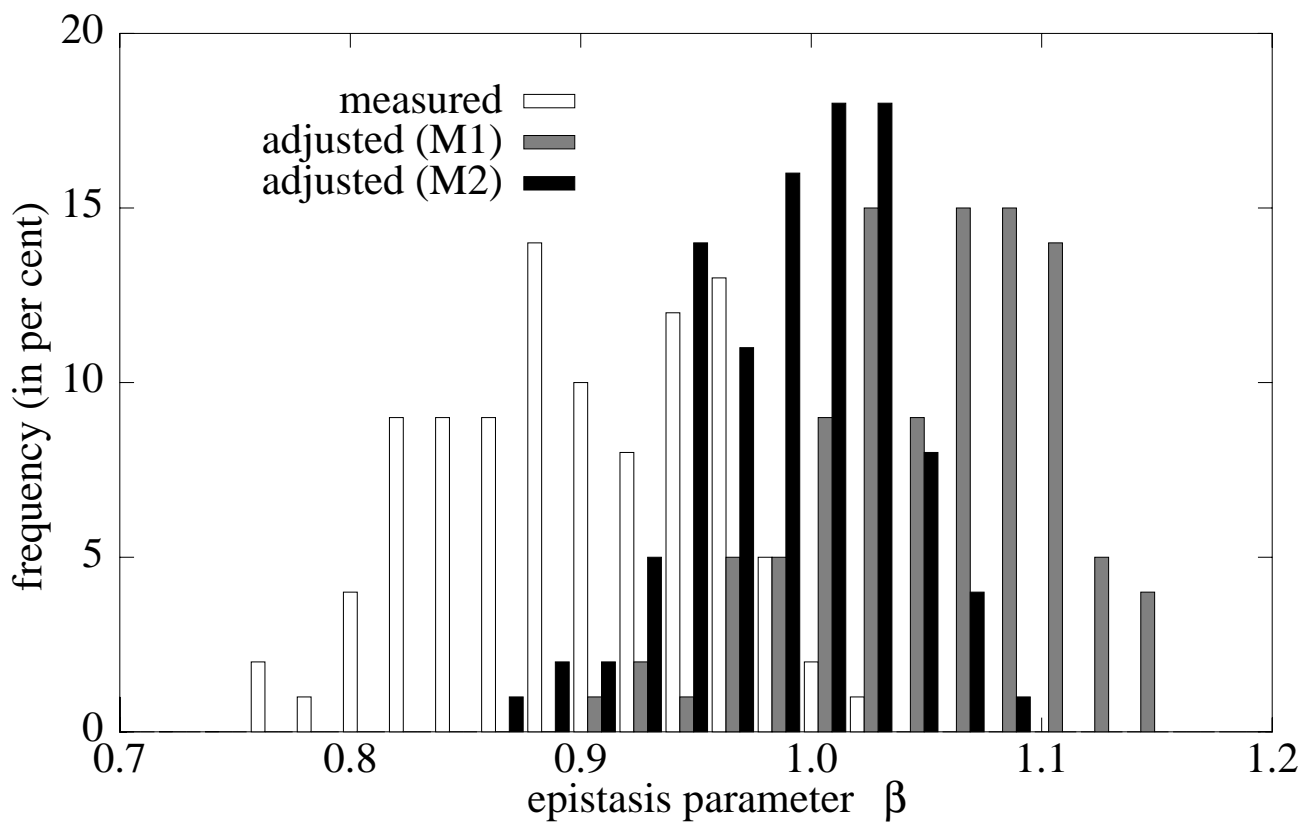

**Figure 7**
Frequency distribution of the epistasis parameter β for measured $w(d)$ and for $w_{neut}(d)$, the latter obtained using two different methods (M1 and M2) to remove the effect of compensatory mutations on $w(d)$. Both methods shift the distribution such that it is roughly centered around 1, in contrast to the strong bias toward antagonism (β < 1) when compensatory mutations are included.

A high abundance of compensatory mutations implies considerable flexibility in the genotype-phenotype map, such that fit phenotypes can be found widely distributed in genotype space. For the case of RNA secondary structure folding, this property of the genotype-phenotype map has been demonstrated analytically using graph theory [41], and it has been verified in extensive numerical simulations [42,43]. Random graph theory predicts that RNA secondary structures can be classified into common structures, which can be found almost anywhere in sequence space, and rare structures. Only a very small number of sequences fold into any given rare structure. The common structures satisfy the shape-covering theorem, which states that within any sphere of a certain radius $r$, at least one sequence can be found that folds into a given common structure. Since we have chosen the reference sequences randomly in our analysis, we can safely assume that the majority of the sequences that we have studied fold into a common structure, and that therefore the shape-covering theorem applies. For sequences of length $L = 76$, we have $r \geq 10$ (Table 3 in Ref. [41]). That means that any common structure can be found at least once within a distance of approximately 10–12 mutations from any arbitrarily chosen sequence. In other words, even in the empty regions of sequence space, at least one out of $10^6$–$10^7$ sequences will be viable. A second prediction from random graph theory is that the common structures should percolate through sequence space. Percolation means that it is possible to find two sequences with almost no sequence identity which fold into the same structure, and which are connected by a path of mutant sequences that also fold into the same structure. However, this percolation property holds only if the exchange of both bases of a pair in paired regions of the secondary structure is considered as a single-mutation event, which runs counter to the standard biological meaning of mutation. Most of these double mutations will be classified as compensatory mutations according to our definition (because each individual mutation of such pairs will most likely destroy the secondary structure), and therefore a single neutral path according to random graph theory will appear as several independent paths linked together by compensatory mutations in our analysis. This fact, combined with the shape-covering theorem which we discussed previously, is probably the source of the large number of compensatory mutations that we observe.

A major difference between the fitness landscapes we have studied here and more natural fitness landscapes is that we restricted fitness values to either one or zero, whereas in nature fitness is certainly not a binary variable. Two sequences that fold into the same structure may bind to a receptor with different efficiencies, and a sequence that folds into a slightly different structure may also bind to the receptor, albeit less efficiently. We chose here binary fitness values for mathematical simplicity and clarity in our analysis. With a continuous fitness scale, apart from neutral and inviable sequences, we would have to keep track of slightly and strongly deleterious as well as advantageous sequences, and would consequently need a much more elaborate mathematical procedure to account for the contributions from these different fitness classes. Nevertheless, it would certainly be desirable to address in a future study the effect of compensatory mutations on epistatic interactions in a system with a continuous fitness scale, for example with a fitness measure similar to the one used by Ancel and Fontana [44] or, alternatively, using digital organisms [2].

The present study does not exclude the possibility that antagonistic epistasis in some systems could be caused by effects other than compensatory mutations. However, our findings do indicate that compensatory mutations can play a major role in producing antagonistic epistasis and thus slowing the decay in mean fitness under random mutation accumulation. Moreover, in the present study, only mutations that re-created exactly the same RNA folding as the wild-type were deemed compensatory, while in nature even a modified secondary structure may have a tertiary structure that is sufficiently close to the wild-type that no fitness loss is incurred [45], which means that compensatory mutations in natural RNA structures may be even more frequent that in our computational model. Compensatory mutations have been reported for rRNA [46], tRNA [47], and mRNA [48] structures, and it is certainly not surprising that we should find a high frequency of compensatory mutations in the present study. What is important and novel in this study is that we have explained the overall excess of antagonistic epistasis by these compensatory mutations. This connection suggests that evidence for one of these phenomena may provide evidence for the other in more complicated systems.

In the remaining paragraphs, we discuss to what extent our results can be applied to systems other than RNA secondary structure folding. We begin with proteins. Several studies of protein folding *in silico* indicate that widespread neutral networks exist, comparable to the situation for RNA sequences [39,49,50]. Moreover, decay functions similar to the function $w(d)$ studied here have been measured *in vitro* [51,52] by making fitness a binary variable (e.g., a protein binds to a receptor or not) and then screening libraries of sequences with varying average numbers of mutations for the relevant property. These experiments found that average fitness initially decayed exponentially with increasing mutations, but the rate of decay tended to decelerate at higher mutation numbers. This deceleration may be attributed to suppressor mutations that stabilize parts of the protein against further mutations [53], which in essence is a form of compensation. For example, the M182T mutation in β-lactamase compensates for locally destabilizing mutations in a loop near the active site [54,55]. Unfortunately, no study has yet investigated explicitly the decay function of proteins that carry such a suppressor mutation. Therefore, it is unclear whether the decelerating decay rates seen by Suzuki et al. [51] and by Daugherty et al. [52] are a consequence of compensatory suppressor mutations or, alternatively, indicate that the reference sequences happened to be located outside a dense cluster of fit sequences.

Using whole bacteria and viruses, several studies have sought to measure the mutational decay function for fitness and the overall direction of epistatic interactions [25–28,56]. Only one of these studies found net synergistic interactions [56]. Several other studies with bacteria and viruses have documented widespread compensatory mutations [57–59], although these studies all used selection to find compensatory mutations rather than directly estimating their frequency. Taken together, these two types of studies are largely consistent with the main conclusions of the present paper, that synergistic interactions do not generally dominate the fitness function and compensatory mutations are common. Moreover, in two of the studies on compensation, it was shown that the compensatory mutations had occurred outside the gene that was debilitated [58,59]. Therefore, the observed patterns of epistasis and compensation do not merely reflect the robustness of individual RNA and protein molecules to multiple mutations (as described above), but must indicate more general features of their genomes.

Of course, it is possible that more complex patterns of gene regulation in eukaryotes [60] might predispose them to synergistic effects between mutations. In fact, studies on fruitflies [30,61] and a unicellular alga [20] (but see [24]) reported net synergistic epistasis, which might reflect fitness landscapes in which peaks are few and far between. But other studies with eukaryotes, including nematodes [27] and fungi [22,29], found no tendency toward synergistic epistasis. Hence, the overall direction of epistasis for fitness, and the possible role of compensatory mutations in pushing toward net antagonistic epistasis among random mutations, will require more study in higher organisms. Finally, Zuckerkandl [62] recently proposed that complex regulatory networks may often evolve via a process of drift decay followed by selection for compensatory mutations. Such an interplay would complicate the overall pattern of directional epistasis relative to regulatory complexity.

## Methods

We predicted RNA secondary structure using the Vienna RNA package, version 1.3.1, with the default set-up [31]. We used 100 randomly chosen RNA sequences of length $L = 76$ as reference sequences, by randomly drawing bases A, G, C, and U from a uniform distribution in which each base has a probability 0.25 of being drawn at any position. For each reference sequence, we determined its secondary structure, and measured the fraction of mutant sequences at Hamming distances $d = 1,2,...,8$ that folded into the same structure, by calculating the secondary structure of up to $10^6$ mutant sequences at each Hamming distance. We used the fraction of neutral mutants at Hamming distance $d$ as the function $w(d)$ for the particular reference sequence. (The fraction of neutral mutants at distance $d$ is equivalent to $w(d)$ under the assumption that mutant sequences which fold into the same structure as the reference sequence have fitness 1, and all other mutants have fitness 0.) In addition, to distinguish neutral from compensatory mutations, we recorded at each Hamming distance $d$, for all viable sequences as well as a random sample of up to 1,000 non-viable sequences, the average number of viable sequences after an *additional* one-step mutation. In the following analyses, we denote the average fraction of viable sequences that are one mutation away from a viable sequence at distance $d$ as $x_{\text{neut}}(d)$. Similarly, we denote as $x_{\text{comp}}(d)$ the average fraction of viable sequences that are one mutation away from a *non-viable* sequence at distance $d$.

As in Ref. [11], we fitted functions $\exp(-\alpha d^\beta)$ to the measured $w(d)$, in order to obtain a two-parameter description of the average effect of single mutations, $\alpha$, and the directional epistasis, $\beta$. We also fitted functions $md + n$ to the average neutrality as a function of $d$, in order to identify whether neutrality was increasing or decreasing with mutational distance $d$.

We used two methods to separate $w(d)$ into contributions from viable sequences that lie on the same neutral network as the reference sequence at $d = 0$ [$w_{\text{neut}}(d)$], and from viable sequences that lie on other neutral networks [$w_{\text{comp}}(d)$].

The first method, M1, uses the information contained in both $x_{\text{neut}}(d)$ and $x_{\text{comp}}(d)$. As explained below, this method tends to overestimate the contribution of $w_{\text{comp}}(d)$. Let $\gamma(d) = w_{\text{neut}}(d) / [w_{\text{neut}}(d) + w_{\text{comp}}(d)]$, which is the fraction of viable sequences belonging to the reference neutral network at distance $d$. Then, $w_{\text{neut}}(d)$ is simply given by

$$w_{\text{neut}}(d+1) = \gamma(d) \, w(d) \, x_{\text{neut}}(d). \quad (2)$$

Estimating $w_{\text{comp}}(d)$ is more complicated. First, we must take into account sequences contributing to $w(d)$ that are not part of the neutral network of the reference sequence. This contribution is given by $[1 - \gamma(d)] w(d) x_{\text{neut}}(d)$. We must then obtain the contribution of additional compensatory mutations at distance $d + 1$. These are given by $1 - w(d)$, the number of non-viable sequences at distance $d$, multiplied by the probability with which these non-viables turn into viable sequences after a single mutation, $x_{\text{comp}}(d)$. Hence, we arrive at

$$w_{\text{comp}}(d + 1) = [1 - \gamma(d)] \, w(d) \, x_{\text{neut}}(d) + [1 - w(d)] \, x_{\text{comp}}(d). \quad (3)$$

With relations (2) and (3), we can recursively calculate $w_{\text{neut}}(d)$ and $w_{\text{comp}}(d)$ for all $d$ for which $w(d)$, $x_{\text{neut}}(d)$ and $x_{\text{comp}}(d)$ are known.

In general, we expect $w(d) \approx w_{\text{neut}}(d) + w_{\text{comp}}(d)$. However, method M1 has the problem that some of the mutations counted in $x_{\text{comp}}(d)$ may actually lead back onto the network of the reference sequence rather than onto a separate neutral network. (While this can be seen as a sort of compensatory mutation, it is not the kind investigated here.) Such sequences would then be counted twice, so that in reality $w(d) < w_{\text{neut}}(d) + w_{\text{comp}}(d)$.

To avoid this double-counting, method M2 uses instead of (3) the expression

$$w_{\text{comp}}(d+1) = w(d+1) - w_{\text{neut}}(d+1). \quad (4)$$

In this case, the quantity $x_{\text{comp}}(d)$ drops out, which insures that $w_{\text{neut}}(d) + w_{\text{comp}}(d)$ is always identical to $w(d)$. M2 does not suffer from double counting viable sequences, but it uses less information than M1 and depends on an additional assumption. In particular, M2 assumes that the viable sequences at distance $d$ that belong to the neutral network of the reference sequence have the same neutrality as those that do not belong to that network. Given these considerations, it seems prudent to use both M1 and M2 rather than rely entirely on either one.

Both methods M1 and M2 are based on the assumption that the fitness landscape can be represented as a tree, and that back mutations can be neglected. With a sequence length of $L = 76$, only one out of 228 possible mutations is a back mutation, and therefore the neglect of back mutations is justified. The representation of the fitness landscape as a tree is potentially more problematic, because there may exist several paths from the reference sequence to a particular sequence at mutational distance $d$, whereas the tree approximation assumes that there is only one such path. For example, consider the wild type aa, with neutral mutants aA, Aa, and AA, Then, at mutational distance one, there are two neutral mutants (aA and Aa), and at distance two, there is only one (AA). The tree approximation counts the mutant at distance two twice, once as a mutant of aA, and once as a mutant of Aa. This double counting means that our methods have the tendency to overestimate $w_{\text{neut}}(d)$. This overestimation, however, is not problematic in the context of the present paper, because – as we emphasized in the results – our main conclusions are based on the large contribution of $w_{\text{comp}}(d)$ to $w(d)$. A correction for the overestimation of $w_{\text{neut}}(d)$ would increase the contribution of $w_{\text{comp}}(d)$, and would therefore make our conclusions even stronger.

## Authors' contributions
COW was responsible for the RNA folding simulations and the data analysis. All authors contributed substantially to the design of the study and the writing of the manuscript. All authors read and approved the final manuscript.

## Acknowledgements
We thank Charles Ofria and Michael D. Stern for stimulating discussions, and Sergey Gavrilets for constructive criticism on an earlier version of this work. This research was supported by the National Science Foundation under Contract No. DEB-9981397. Part of this work was carried out at the Jet Propulsion Laboratory, California Institute of Technology, under a contract with the National Aeronautics and Space Administration.